# In Situ X-Ray Radiography and Tomography Observations of the Solidification of Alumina Particles Suspensions

## Part II: Steady State

Sylvain Deville[1*], Eric Maire[2], Audrey Lasalle[1], Agnès Bogner[2], Catherine Gauthier[2], Jérôme Leloup[1], C. Guizard[1]

[1] Laboratory of Synthesis and Functionnalisation of Ceramics, UMR3080 CNRS/Saint-Gobain CREE, Cavaillon, France

[2] Université de Lyon, INSA-Lyon, MATEIS CNRS UMR5510, 7 avenue Jean Capelle, F-69621 Villeurbanne, France

## Abstract

This paper investigates the behaviour of colloidal suspensions of alumina particles during directional solidification, by in situ high-resolution observations using X-ray radiography and tomography. This second part is focussed on the evolution of ice crystals during steady state growth (in terms of interface velocity) and on the particles redistribution taking place in this regime. In particular, it is shown that diffusion cannot determine the concentration profile and the particles redistribution in this regime of interface velocities (20-40 microns/s); constitutional supercooling arguments cannot be invoked to interpret particles redistribution. Particles are redistributed by a direct interaction with the moving solidification interface. Several parameters controlling the particles redistribution were identified, namely the interface velocity, the particle size, the shape of the ice crystals and the orientation relationships between the crystals and the temperature gradient.

## 1. Introduction

The behaviour of a propagating solidification interface in a suspension of particles is of major interest for its implications in a very large number of theoretical and practical issues. If the physical mechanisms controlling the interactions are relatively well understood for single large particles, the solidification behaviour of colloidal suspensions is a matter of great interest, but highly challenging. Conclusions derived from single large particles experiments [1-6] can hardly be extrapolated to smaller particles (<1 microns) where Brownian motion is dominating and segregation effects are negligible. The analysis is further complicated by the necessity to take into account the various interactions between particles, which could be of different natures: electrostatic, Van der Waals, steric, etc… Additional deviations from the ideal situation, which could be modelled, such as the distribution of particle size, their surface state, charge and roughness, could have a major influence over the general behaviour and stability of the system but are difficult to take into account theoretically. We focus here on the behaviour of an interface propagating in a steady state (in terms of velocity), in a concentrated suspension of colloidal particles. A convenient approach to tackle the problem is to apply the formalism developed for the solidification of alloys, considering that particles can diffuse like atoms (although with much slower diffusion kinetics), with a dual-phase microstructure formation, one phase being the pure ice and the other one the concentrated particles entrapped in ice. If some common characteristics between these two systems are encountered, such as the relationship between the structural wavelength and the interface velocity [7], it is nevertheless necessary to incorporate the specificities associated to particles in the analysis. Such an effort is under way, under the so-called "colloidal alloys" designation [8-11]. Experimental results gathered so far concentrated on low interface velocities systems (<1 micron/s), which are hardly relevant to the conditions encountered in the materials processing route called freeze-casting. Conclusions were nonetheless derived for these systems, with a particular importance given to the diffusion of particles, which in turn might create conditions of constitutional supercooling, with consequences on crystals growth kinetics and particles redistribution.

---

[*] Corresponding author : sylvain.deville@saint-gobain.com

It has previously been shown that during steady state, particles are repelled by the propagating interface and are concentrated between the growing crystals. Particles concentration increases up to the breakthrough concentration, where the interface moves into the interparticle space [12]. This situation corresponds to the propagation of a solidification interface in a porous medium and has been greatly investigated, both theoretically and experimentally [13, 14].

A common conclusion to all the existing studies of such systems is the existence of a critical interface velocity (or correspondingly a critical particle size for a given velocity) above which particles are engulfed by the moving interface before the maximum particle concentration can be reached.

The steady state behaviour of a system is investigated here, where the particle size is relatively large (0.2-3.4 microns) in comparison to the typical colloidal systems (<0.1 microns), but smaller than previous studies (>1 microns) and the interface velocity is relatively large too (>20 microns/s). This corresponds to what is encountered during the processing of porous materials by the so-called freeze-casting process [15, 16], or solvent-induced phase separation. Under these conditions, diffusion cannot take place, the interface velocity being too high, and the particles redistribution is obtained by the direct interaction with the solidifying interface, which considerably simplify the analysis. The growth kinetics were directly measured (although the in-plane growth anisotropy between the a and c axis was not accessible), and several parameters controlling the redistribution of particles were identified, in agreement to those identified in the first companion paper.

## 2. Experimental methods

All the experimental methods have been described in the first companion paper, "In situ X-ray radiography and tomography observations of the solidification of alumina suspensions, Part I: Initial Stages". In the present part, additional powders were also used; their characteristics are given in table 1. The steady state regime observations were performed in the area 4 to 6 mm from the nucleation surface.

## 3. Results

### 3.1 Steady-state solidification

Images of the steady state solidification regime of the 1.3 microns particles suspension, obtained by radiography, are shown in Figure 1. The displacement of the interface with time can be clearly observed in the pictures. Using these sequences, it was possible to track the position of the interface with time, to obtain a value for the interface velocity; data are plotted in figure 2. The interface exhibits a smooth displacement, with a small but noticeable decrease of the interface velocity with time, which can be explained through the imposed temperature conditions (linear cooling ramp). The average interface velocity was around 39 microns/s in this area of the sample. Two peaks in the displacement and velocity data can be noticed, respectively, called A and B in figure 2. Similar to what was observed in figure 5 of part I, these peaks indeed correspond to instabilities but will be discussed in a separate paper.

A full tomography acquisition of the same part of the sample was performed once solidification was completed, providing after reconstruction a three-dimensional picture of the ice and particles phases, as shown in figure 3 and 4. The vertical cross-section reveals continuous crystals, running from the bottom to the top of the area. Under these solidification conditions, crystals are growing continuously. The corresponding horizontal cross-sections shown in figure 3 b to e reveal the arrangement of the ice crystals in the direction perpendicular to the solidification direction. Crystals are lamellar, as previously discussed, but arranged in domains with similar orientations (figure 3 f to i). The orientation of each domain can be related to the original nucleation conditions, as discussed in the previous paper. The arrangement of the domains is retained over large distances, as seen on the sequence of cross-sections. The displacement of the domains boundaries can be partially explained by the small tilt of the crystals (a few degrees, visible in figure 3a).

Quantitative information can be extracted from these images, so that the particle rich phase fraction (as discussed previously) can be plotted versus the position of the interface (figure 5), along with the corresponding interface velocity obtained through the radiography data. The interface velocity is slowly decreasing, while the particle fraction is constant, with variations within the measurement error range (estimated at 0.5%). Hence the particle fraction is not dependent on the interface velocity, under these conditions.

Using the horizontal cross-sections, the structural wavelength of the frozen structure can be measured; this parameter corresponds to the average *crystals + entrapped particles* length measured perpendicular to the c-axis of the crystals, as measured before [17]. The structural wavelength is varying with the interface velocity (figure 6), from 41 to 35 microns, as shown before in such systems [7, 12], although we have here a direct and local measurement of the relationship, which is not the case when measured after sublimation and sintering, with the corresponding shrinkage accompanying densification.

## 3.2 High-resolution observations

High-resolution observations were performed on the same systems (0.4 microns particles) with the same temperature conditions, to provide direct observations of the crystals growth and particles displacement. Such sequence is shown in figure 7 and movie 1. The first observation is the confirmation of the lamellar morphology of the ice crystals, as reported previously. The liquid-solid transformation of water into ice is accompanied by a 4 vol.% increase. This increase leads to a displacement of the suspension placed above the ice crystals. This displacement can clearly be observed in movie 1, with a vertical translation of the suspension as the solid-liquid interface is progressing.

The tip of the ice crystals is particularly visible, and a zone of concentrated particles can be seen around the interface (dark area). A detail is shown in figure 8, where the tip of the ice crystals can be seen more clearly, just emerging from the concentrated particles zone. It is quite clear that under these conditions the particles are not concentrated ahead of the tip of the crystals, but in-between the crystals.

Similar observations with larger particles (1.3 microns particles) were performed, to investigate the influence of particle size. The sequence is shown in figure 9 and movie 2. A similar behaviour is observed, with one noticeable difference: the presence of a concentrated particles zone ahead of the interface. The crystals tips are not emerging from this concentrated zone. In addition, the height of this zone is increasing as the interface is advancing. Segregation of the particles can also be observed in movie 2, with particles progressively falling towards the bottom of the suspension; which is not surprising considering the relatively larger size of the particles.

The position of the interface and its velocity can be measured from these observations; their values are plotted in figure 10, for the two types of particles (0.4 and 1.3 microns). As shown before (figure 4 and 5), the interface velocity is slowly decreasing with time, from 24 to, respectively, 17 and 12 microns/s for small and large particles (faster decrease for larger particles).

## 4. Discussion

### 4.1 Crystals growth

Several hypotheses were made concerning the morphology of the ice crystals growing in colloidal suspensions of particles, mostly based on the anisotropy of growth kinetics. From observations of materials obtained through solidification of colloidal suspension, the observed lamellar porous morphology seems to indicate a lamellar shape for the ice crystals, in good agreement with the strong anisotropy of growth kinetics of the a- and c-axis of hexagonal ice [18, 19]. The particles arrangement can nevertheless be expected to be affected by the high-temperature densification step (sintering). Such lamellar morphology can be directly observed here, in particular on the high resolution 2D observations in figure 7 and in 3D in figure 11, confirming the previous hypotheses. Additional growth behaviours usually associated with dendritic growth, such as tip splitting leading to twin crystals and protruding dendrites turning into primary crystal can be observed in figure 4a (arrows). Such instabilities seem to rarely yield stable dendrites here. By the time such protruding dendrites turn into stable dendrites growing along the z-axis, the surrounding z-crystals have already grown and induced concentration of particles ahead of the newly created dendrites. The remaining space available for their growth is therefore very limited; they rapidly come to a halt, either by stopping halfway in between the lamellar crystals (Figure 12 a), or by reaching the next adjacent crystals (Figure 12 b).

From the radiography sequences, where the evolution of the crystals and particles redistribution below the interface is visible, the growth kinetics along the a- and c- axis (Va$_z$ and Va$_{xy}$) can also be estimated. The various interface velocities are schematically illustrated in Figure 13. Three different velocities exist in this situation, as a consequence of the crystal growth anisotropy and the thermal conditions of directional solidification: the Vc$_{xy}$ and Va$_{xy}$ respectively correspond to the growth velocities along the c and x axis, in the xy plane, while Va$_z$ corresponds to the growth velocity along the a axis perpendicular to the xy plane. Combined with the crystals dimensions measured on cross-sections obtained through tomography, the c-axis growth rate Vc$_{xy}$ can be estimated around 14 microns/s, for an a-axis growth rate Va$_z$ of 18 microns/s in the solidification direction, with large particles (1.3 microns). The c-axis and a-axis growth rates Vc$_{xy}$ and Va$_z$ were estimated, respectively, around 14 microns/s and 20 microns/s with small particles. No noticeable effect of the particle size can be observed here. The a- and c-axis growth kinetics are very similar, and more similar than expected. What we measure here is not the actual anisotropy of growth kinetics (defined as Va$_{xy}$/Vc$_{xy}$), as the growth rate measured in the z-direction is dictated by the temperature gradient, which is not the case for the growth in the x- or y-direction (figure 13). To access the anisotropy of growth kinetics, one should measure the growth rate for the a- and c-axis within the xy-plane (Va$_{xy}$ and Vc$_{xy}$), where the temperature conditions are homogeneous, so that the measured growth rate are not dependent on the temperature gradient. Such measurement cannot be performed with these types of experiments, until full tomography acquisitions can be performed within a much shorter time to obtain in situ three-dimensional evolution of the crystals growth.

## 4.2 Particles redistribution

From the various observations shown here, it appears that redistribution of particles in front of the advancing interface is dependent on several variables. Under certain conditions, no particles are found ahead of the interface, while under other conditions, a layer of concentrated particles is clearly visible ahead of the interface. Several parameters have been identified here, and are discussed below.

- Particles redistribution depends on the interface velocity. This dependency seems rather obvious, and can be understood when interpreting the system in terms of colloidal particles and diffusion. However, when the interface is moving fast enough, particles cannot migrate by Brownian diffusion, so that the problem cannot be treated in terms of diffusion. Particles are directly repelled by the moving solidification front. For moderate to slow velocities (1 microns/s), enough time is available for the particles to migrate by Brownian diffusion, so that a layer of concentrated particles can build up ahead of the interface. This situation has already been deeply investigated [9, 10]. In such case, the concentration of particles ahead of the interface can potentially lead to a situation of constitutional supercooling, where ice crystals will be able to nucleate within the concentrated zone, leading to a different morphology of particles arrangement in the solidified body. This was observed here for very slow growth rate, when the crystals' growth was just about to stop (figure 14). In situ tomography acquisitions, only possible for very slow growth rates, reveal the nucleation of crystals within the concentrated particles zone. This situation could be detrimental when used for processing of materials, leading to large size defects and discontinuities in the final structure. Yet, such conditions are never used, for practical reasons. The structural wavelength of the structures obtained with such slow interface velocities are not of interest so far. Under the temperature conditions typically used, it is quite clear that diffusion is never taking place, and thus the diffusion approach cannot be used to determine the concentration profile and particles redistribution behaviour in this case.

- Particles redistribution depends on particle size, or, more precisely, on the ratio of particle size and intercrystals spacing. Although the relationship between particle size and diffusion is quite clear (see for example Fig. 3 in reference [11]), additional effects are associated to the particle size, under conditions where no diffusion is possible. As shown in figure 7 and 9, particles redistribution is more difficult with larger particles. A zone of concentrated particles builds up when larger particles are used, a situation that cannot be explained through diffusion since this mechanism is less likely to occur as particle size is increasing. It is proposed here that this concentration ahead of the interface is related to the difficulty of obtaining the maximum packing density of particles between the crystals when the particle size falls within the same range of order as the crystals separation distance, as mentioned and schematically illustrated in figure 8 in the companion paper. The presence of a concentrated particles zone can affect the growth of the ice crystals, their growth kinetics was found to slow down faster when concentrated particles were present (figure 10). Again, this could be detrimental for the final materials, as this interface deceleration is associated with an increase of the structural wavelength. The resulting materials will be less homogeneous than those obtained with smaller particles.

- Particles redistribution depends on the shape and the orientation relationship between the crystals and the temperature gradient direction. This has been shown in the companion paper (part I: Initial instants), and is confirmed here. Under steady state conditions, the crystals are growing along the temperature gradient direction; the measured particle rich phase fraction in this direction is constant.

In addition, local instabilities can be observed, leading to disruption of the crystals continuity along the temperature gradient direction. Disruptions are shown in figure 15a. They are detrimental for the materials processed through this technique, as they are not healed with the densification stage, so that they turn into crack-like defects in the sintered pieces (figure 15b). Mechanical properties of these highly cracked structures are dramatically lower than their defect-free counterparts. The underlying reason for this behaviour is not clear at this time. It is dependent on the formulation, as a proper one can provide defect-free sintered bodies with excellent mechanical properties [20].

## 4.4 Analogies and disparities with alloys solidification and solidification in porous media

Previous experimental and theoretical investigations regarding the solidification of suspensions have revealed two extreme cases, depending on the particle size: the problem can either be treated using Fick's law, which is similar to what is done in the case of the solidification of alloys, or using Darcy's law, considering the propagation of a solidifying interface in a porous media, when particles of larger size are used. For alloys, transport properties exhibit a linear concentration dependence, a characteristic that greatly simplify the analysis. Further improvements were performed when the interactions between particles were taken into account, to characterize the behaviour of the system from a colloidal suspension point of view. In this situation, particles redistribution can still occur by diffusion, but the system is characterized by a strongly nonlinear concentration dependence of the transport properties, and in particular a nonlinear freezing temperature curve [10]. The interface velocities investigated were nevertheless much lower (0.1-0.8 microns/s) that the ones observed here (20-40 microns/s). The other extreme situation is the propagation of a solidifying interface in a porous media. In such case, no particles redistribution can take place and obviously no diffusion can occur. This extreme is too different from the current situation when particles redistribution by the moving interface begins, considering the range of particle size and concentration used here, with the major consequence that particles redistribution induced by the moving interface plays a major role in the structural arrangement of the frozen body. However, when the lateral growth of crystals is large enough, particles concentration increases up to the maximum possible concentration, corresponding to a situation where the movement of individual particles is hardly possible, typical particles concentration at this point being close to the maximum packing. In this situation, the particles are engulfed by the propagation of the interface in the interparticle space, without movements of particles; Darcy's law is in this case relevant to describe the behaviour of the system.

The mechanisms involved here (before particles concentration get close to its maximum) are similar to the solidification of alloys, although several major differences are observed. The particle size are in a range (0.2-3.4 microns) where diffusion could occur, but the interface velocities applied here are too fast for diffusion to occur. No particles redistribution can occur by diffusion, as observed directly on high resolution radiography results. Therefore, the influence of concentration dependent properties have less influence than for colloidal alloys investigated both theoretically and experimentally. The first consequence is that no situation of constitutional supercooling can occur. The second consequence is that the redistribution of particles in directions parallel and perpendicular to the interface displacement direction is not dictated by diffusion, but rather by the orientation of the growing crystals and the efficiency of particles packing when the particles concentration is increasing between the moving interfaces. As a consequence, one should refer to this situation as a metastable steady state, dictated by the applied temperature conditions (cooling rate). Given more time, particles would diffuse away from the interface and reach an equilibrium situation, as observed experimentally [10, 21, 22]. Providing that the crystals orientation is constant and the intercrystals distance large enough (compare to the particle size), particles redistribution occurs homogeneously in the xy-plane (with a constant packing efficiency) and does not occur in the z-direction. The fraction of entrapped particles is therefore constant, resulting in homogeneous materials after removal of the solvent.

## 5. Conclusions

Based on in situ experiments by X-ray radiography and tomography of the controlled solidification of concentrated alumina particles suspensions, the following conclusions can be drawn, regarding the steady state regime (in terms of velocity):

- Due to the relatively rapid interface velocities (20-40 microns/s), particles redistribution cannot occur by diffusion. Situation of constitutional supercooling cannot be involved here.

- Particles redistribution is obtained through direct interactions with the solidification front, particles being pushed by the moving interface. The critical parameters to be considered here are therefore the orientation of the growing crystals and the efficiency of particles packing when the particles concentration increases between the moving interfaces, efficiency directly related to the particle size and intercrystals spacing.

## Acknowledgements

We acknowledge the European Synchrotron Radiation Facility for provision of synchrotron radiation beam time and we would like to thank Elodie Boller and Jean-Paul Valade for their irreplaceable assistance in using beamline ID19. Financial support was provided by the National Research Agency (ANR), project NACRE in the non-thematic BLANC programme.

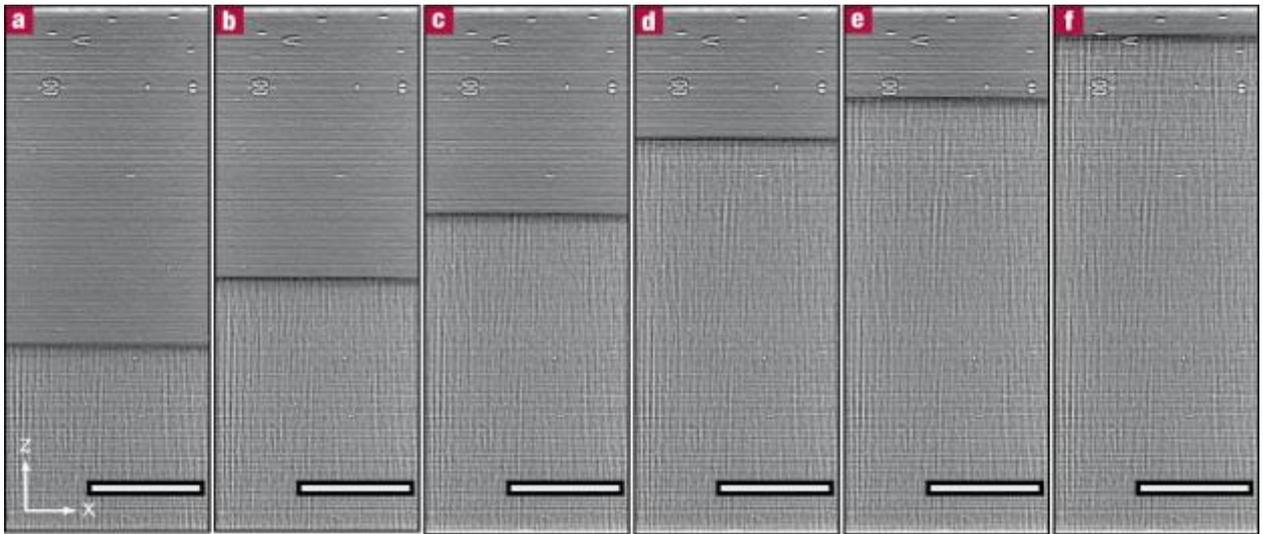

Figure 1: Solidification sequence in the steady state, radiography, 1.3 microns particles. Time picture was taken at: a=$t_0$, b= $t_0$+6s, c= $t_0$+13.7s, d= $t_0$+22.3s, e= $t_0$+27.3s, f= $t_0$+34.7s. Scale bar: 500 microns.

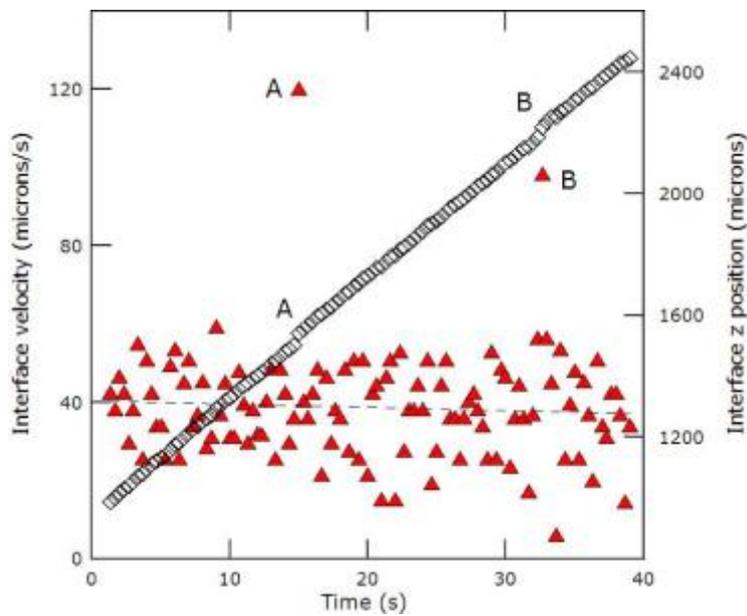

Figure 2: Interface location (◊) and velocity (▲, dashed line: linear fit) vs. time, steady state, 1.3 microns particles. The average velocity of the interface is 39 microns/s. The z=0 position corresponds to the position of the interface in Fig. 1a (time $t_0$). The velocities peaks noted A and B are significant, but will be discussed in a separate paper.

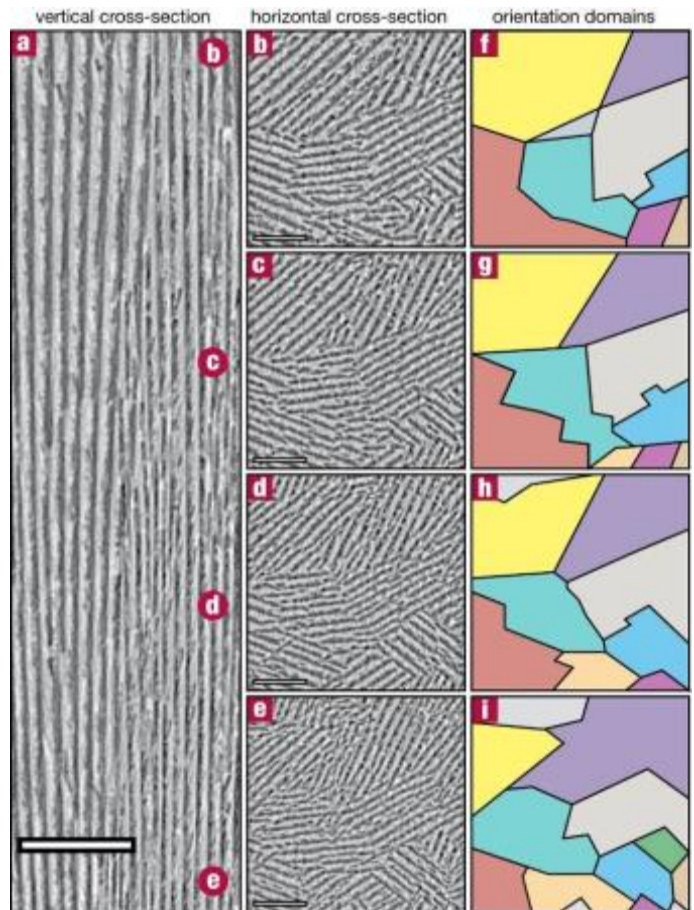

Figure 3: Reconstruction from tomography acquisition, 1.3 microns particles. Vertical (a) and horizontal (b-f) cross-sections of the frozen arrangement. The corresponding orientation domains, corresponding to domains of long range order of the crystals alignment, are given in (f)-(g). Scale bars : *a* : 300 microns, b-e : 150 microns.

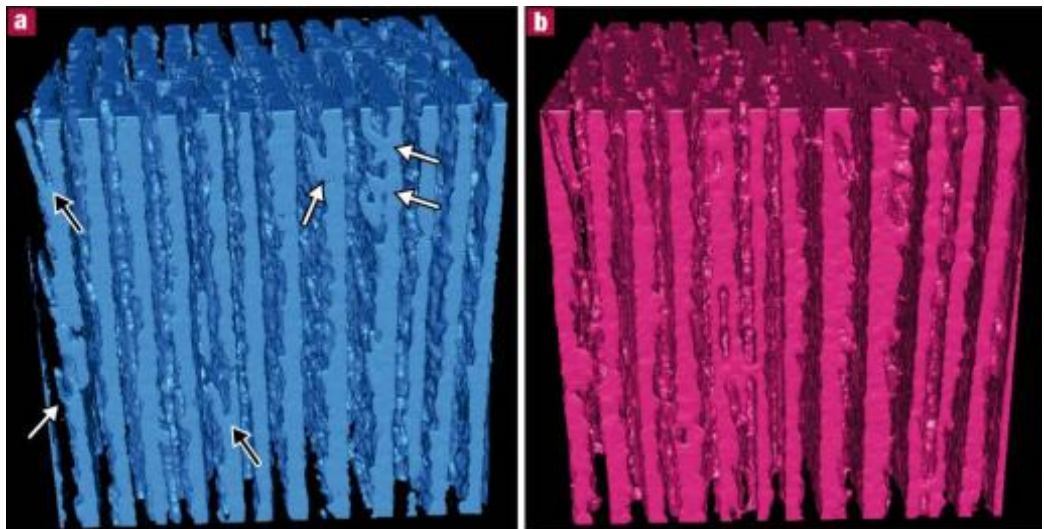

Figure 4: Three-dimensional representation of a block of lamellar ice crystals (left) and corresponding particles entrapped (right). Block dimension: 360x360x360 microns$^3$, 1.3 microns particles. White arrows indicate the protruding dendrites, which can eventually turns into lamellar crystals, black arrows indicate twin dendrites, probably resulting from tip splitting.

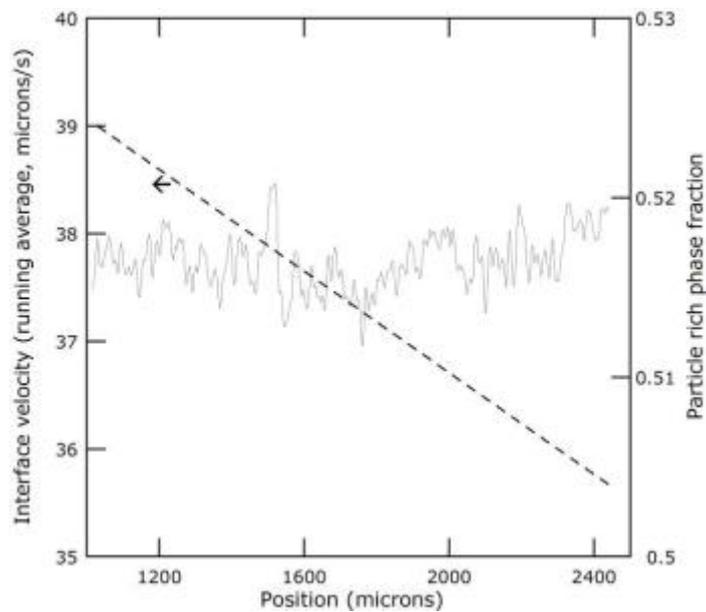

Figure 5: Interface velocity (fit, dashed line) and particle fraction (continuous line) vs. position of the interface, steady state, 1.3 microns particles. The z=0 position corresponds to the bottom of the picture in Fig. 1a. No significant variations of the particles fraction are observed.

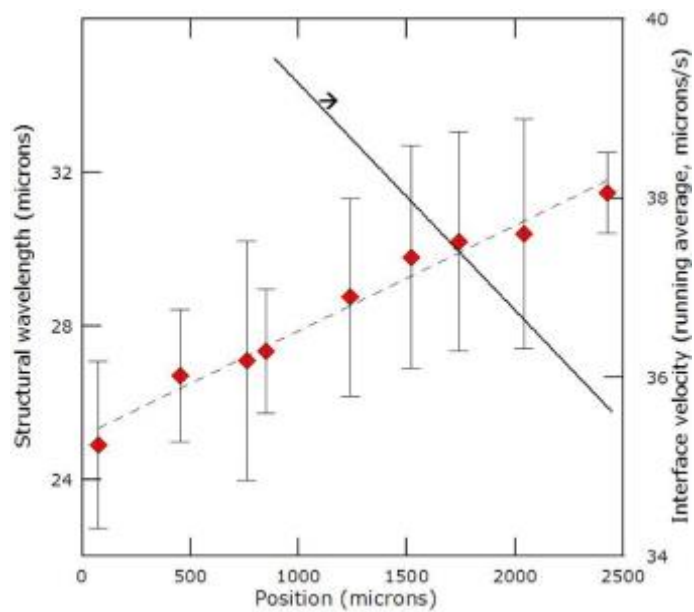

Fig. 6: Structural wavelength (◊) and interface velocity (fit)(▲) vs. position, as measured on reconstructed cross-sections. 1.3 microns particles.

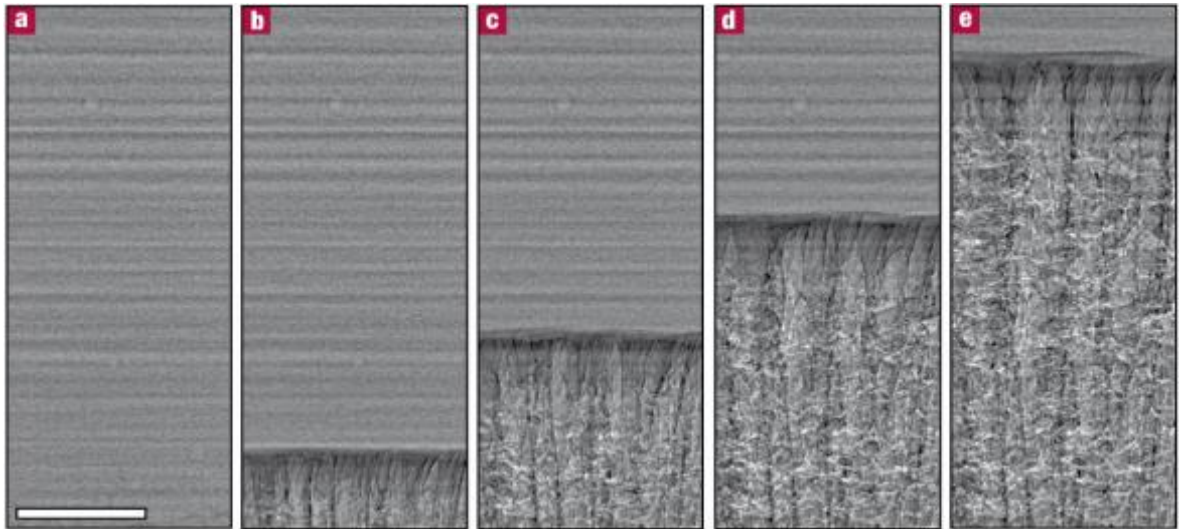

Figure 7: Solidification sequence in the steady state, high resolution radiography, 0.4 microns particles. Time picture was taken: a=$t_0$, b= $t_0$+4s, c= $t_0$+9.7s, d= $t_0$+16s, e= $t_0$+24.7s. Scale bar: 150 microns.

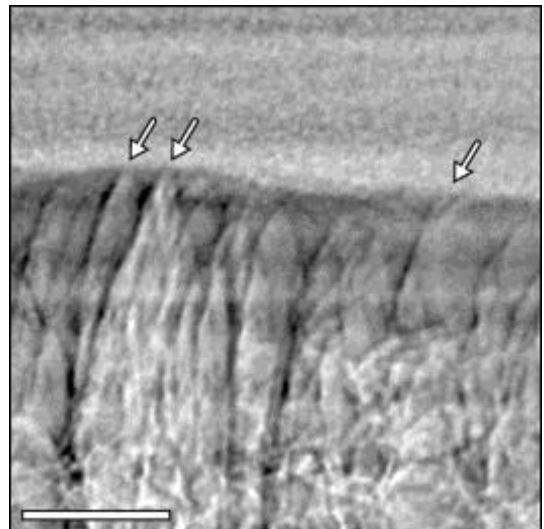

Figure 8: Detail of the solidification sequence in the steady state (Figure 7), high resolution radiography, 0.4 microns particles. Details of the dendrites tip. No visible concentration of particles ahead of the dendrites tip is observed. Interface velocity: 20 microns/s. Scale bar: 50 microns. Arrows indicate the dendrites tip. The solidification front is moving vertically, bottom to top.

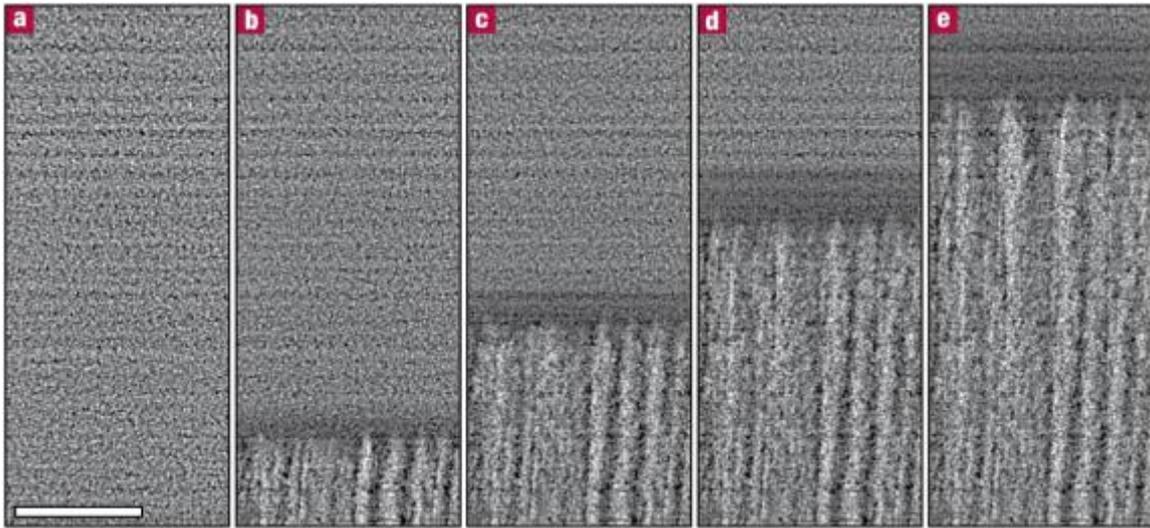

Figure 9: Solidification sequence in the steady state, high resolution, 3.4 microns particles. Time picture was taken: a=$t_0$, b= $t_0$+4s, c= $t_0$+9.7s, d= $t_0$+16s, e= $t_0$+24.7s. Scale bar: 150 microns.

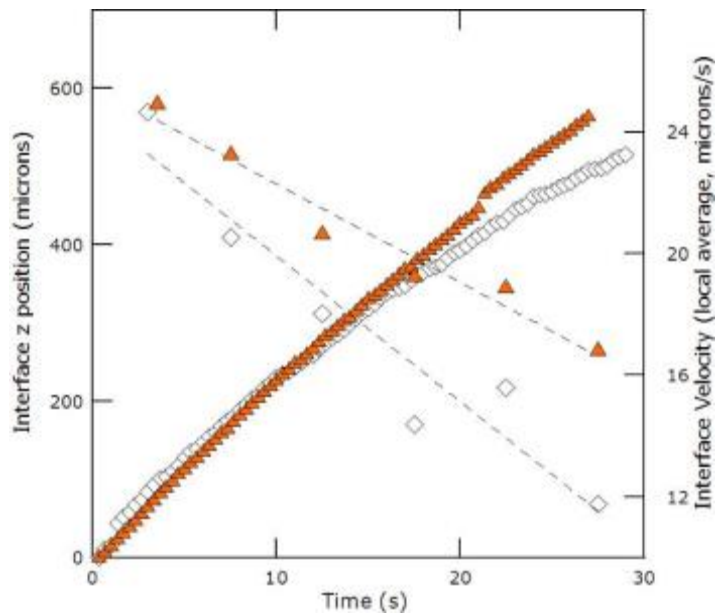

Figure 10: interface position and velocity from high resolution observations, for 0.4 microns (▲) and 3.4 microns (◊) particles.

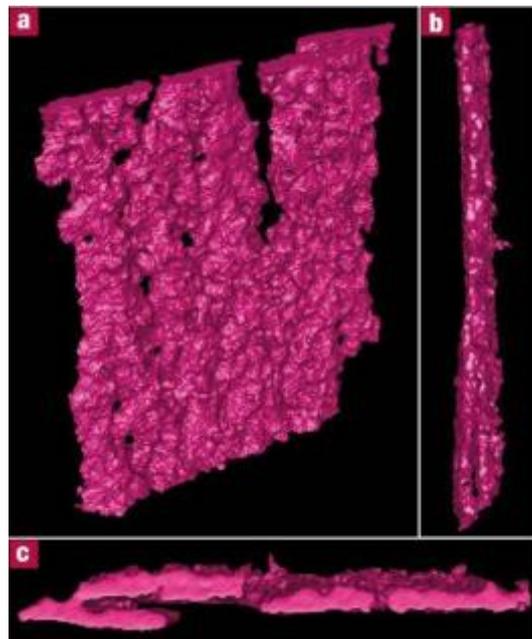

Figure 11: Three-dimensional representation of an isolated lamellar crystal, 1.3 microns particles. Perspective (a), side (b) and top view (c). Crystal height represented: 360 microns. The surface of the crystal is not faceted.

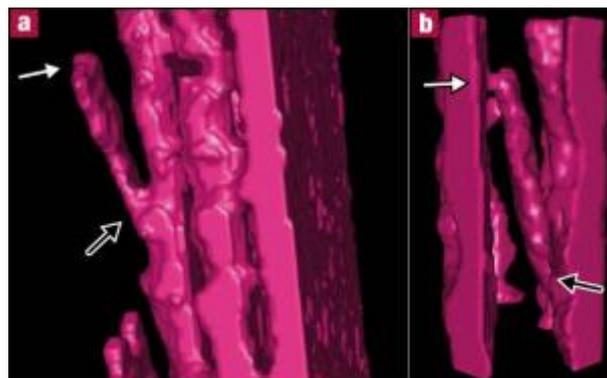

Figure 12: Growth termination of dendrites, three-dimensional representation. Black arrows indicate the origin of the dendrites and white arrows their termination. The dendrites can either stop in between the crystals (a), or reach the next adjacent crystal (b).

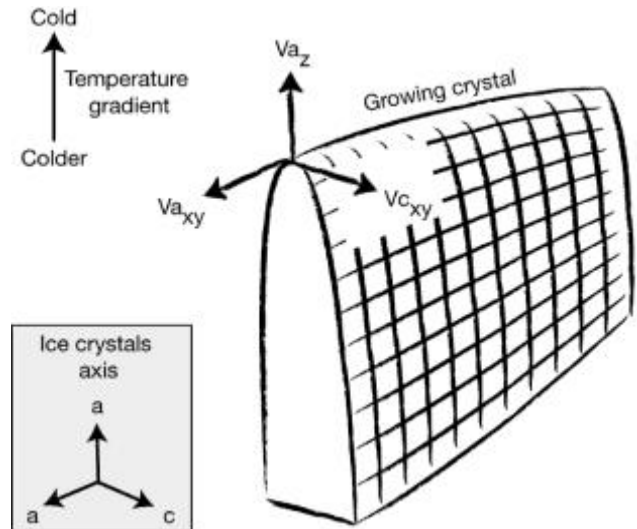

Figure 13: Growth velocities and crystal orientation. Ice, under these conditions of temperature and pressure, is hexagonal, with strong anisotropies of interface kinetics between the a and c axis, resulting in platelets-like crystals.

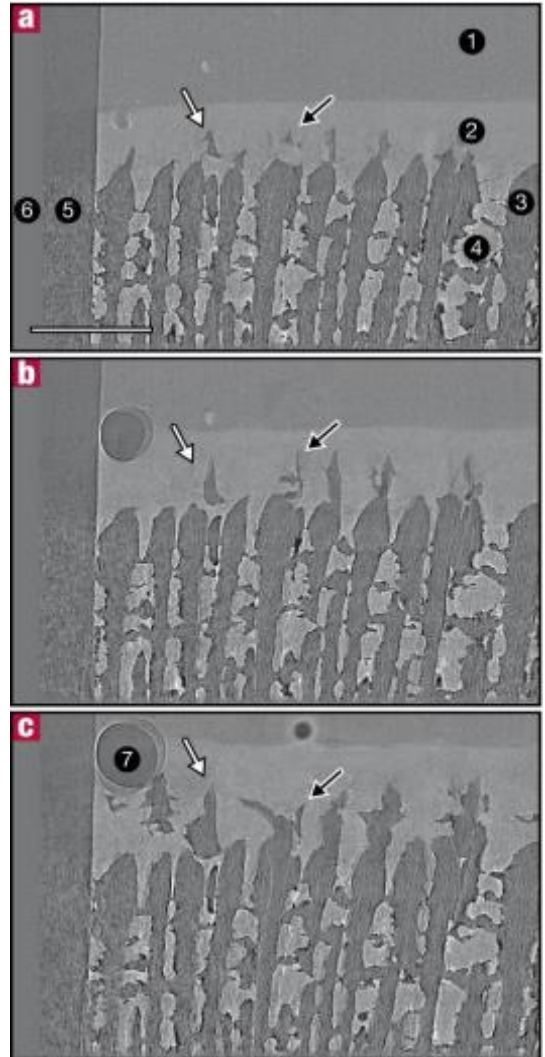

Figure 14: Solidification sequence, in situ tomography, 0.4 microns particles. Evidences of constitutional supercooling. The solidification front is moving vertically, bottom to top. Scale bar: 500 microns. Legend: 1: unfrozen suspension, 2: concentrated particles ahead of the interface, 3: ice crystals, 4: concentrated particles surrounded by ice 5: polymer mold, 6: outside background, 7: bubble. Approximate time: a: $t_0$, b: $t_0+30s$, c: $t_0+60s$.

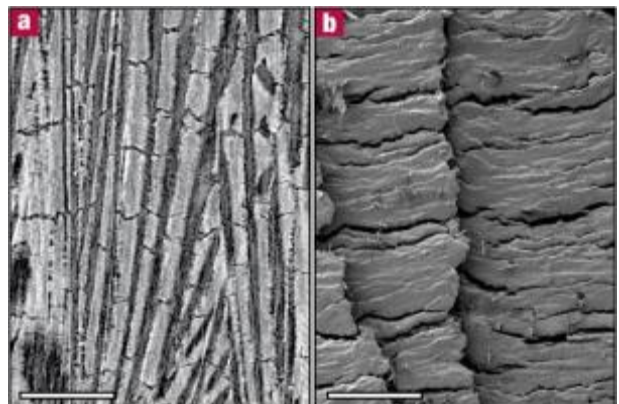

Figure 15: Evidence of local instabilities in the frozen structure (a), leading to large crack-like defects in the final structure (b). Scale bars: 200 microns. a: tomography picture, b: SEM picture after sublimation and sintering.

| Manufacturer | Product name | $D_{50}$ (microns) | SSA ($m^2/g$) |
|---|---|---|---|
| Taimei | TM DAR | 0.2 | 14.5 |
| Ceralox | SPA 0.5 | 0.4 | 8 |
| Almatis | CT1200SG | 1.3 | 3.7 |
| Almatis | CT800SG | 3.4 | 1 |

Table 1: Characteristics of powders used in this study